\documentclass[cits]{PoS}

\title{NNLO predictions for event shapes and jet rates in electron-positron annihilation}

\ShortTitle{NNLO predictions for event shapes and jet rates in electron-positron annihilation}

\author{\speaker{Stefan Weinzierl}\\
        University of Mainz\\
        E-mail: \email{stefanw@thep.physik.uni-mainz.de}}

\abstract{
The strong coupling constant is a fundamental parameter of nature.
It can be extracted from experiments measuring three-jet events in
electron-positron annihilation.
For this extraction precise theoretical calculations for jet rates
and event shapes are needed.
In this talk I will discuss the NNLO calculation for
these observables.
          }

\FullConference{RADCOR 2009 - 9th International Symposium on Radiative Corrections (Applications of Quantum Field Theory to
Phenomenology) \\
                 October 25-30 2009\\
                 Ascona, Switzerland}

\begin{document}

\section{The calculation}

The process $e^+ e^- \rightarrow 3 \; \mbox{jets}$ is of particular interest for the measurement of the
strong coupling $\alpha_s$. 
Three-jet events are well suited for this task because the leading term in a perturbative calculation 
of three-jet observables is already proportional to the strong coupling.
For a precise extraction of the strong coupling one needs in addition to a precise measurement of three-jet
observables in the experiment a precise prediction 
for this process from theory. This implies the calculation of higher order corrections.
The process $e^+ e^- \rightarrow 3 \; \mbox{jets}$ has been been calculated recently 
at next-to-next-to-leading order (NNLO) in 
QCD \cite{GehrmannDeRidder:2007hr,Weinzierl:2008iv}.
The master formula for the calculation of a three-jet observable at an 
electron-positron collider is
\begin{eqnarray}
\langle {\cal O} \rangle & = & \frac{1}{8 s}
             \sum\limits_{n \ge 3}
             \int d\phi_{n}
             {\cal O}_n\left(p_1,...,p_n,q_1,q_2\right)
             \sum\limits_{helicity} 
             \left| {\cal A}_{n} \right|^2,
\end{eqnarray}
where $q_1$ and $q_2$ are the momenta of the initial-state particles and $1/(8s)$ corresponds to the flux factor 
and the average over the spins of the initial state particles.
The observable has to be infrared safe, in particular this implies that in single and double
unresolved limits we must have
\begin{eqnarray}
{\cal O}_{4}(p_1,...,p_{4},q_1,q_2) & \rightarrow & {\cal O}_3(p_1',...,p_3',q_1,q_2)
 \;\;\;\;\;\;\mbox{for single unresolved limits},
 \nonumber \\
{\cal O}_{5}(p_1,...,p_{5},q_1,q_2) & \rightarrow & {\cal O}_3(p_1',...,p_3',q_1,q_2)
 \;\;\;\;\;\;\mbox{for double unresolved limits}.
\end{eqnarray}
${\cal A}_n$ is the amplitude with $n$ final-state partons.
At NNLO we need the following perturbative expansions of the amplitudes:
\begin{eqnarray}
& & 
 \left| {\cal A}_3 \right|^2
 =  
   \left| {\cal A}_3^{(0)} \right|^2
 + 
   2 \; \mbox{Re}\;
   \left(
             \left. {\cal A}_3^{(0)} \right.^\ast {\cal A}_3^{(1)} 
       \right)
 + 
   2 \; \mbox{Re}\;
   \left(
             \left. {\cal A}_3^{(0)} \right.^\ast {\cal A}_3^{(2)} 
       \right)
           + \left| {\cal A}_3^{(1)} \right|^2,
 \nonumber \\
& &
  \left| {\cal A}_4 \right|^2
 =  
   \left| {\cal A}_4^{(0)} \right|^2
 + 
   2 \; \mbox{Re}\;
   \left(
          \left. {\cal A}_4^{(0)} \right.^\ast {\cal A}_4^{(1)} 
 \right),
 \nonumber \\ 
& &
  \left| {\cal A}_5 \right|^2
 = 
   \left| {\cal A}_5^{(0)} \right|^2.
\end{eqnarray}
Here ${\cal A}_n^{(l)}$ denotes an amplitude with $n$ final-state partons and $l$ loops.
We can rewrite symbolically
the LO, NLO and NNLO contribution as 
\begin{eqnarray}
\label{weinzierl_nnlo_def_LO_NLO_NNLO}
\langle {\cal O} \rangle^{LO} & = & \int {\cal O}_{3} \; d\sigma_{3}^{(0)},
 \nonumber \\
\langle {\cal O} \rangle^{NLO} & = & \int {\cal O}_{4} \; d\sigma_{4}^{(0)} + \int {\cal O}_{3} \; d\sigma_{3}^{(1)},
 \nonumber \\
\langle {\cal O} \rangle^{NNLO} & = & \int {\cal O}_{5} \; d\sigma_{5}^{(0)} 
                   + \int {\cal O}_{4} \; d\sigma_{4}^{(1)} 
                   + \int {\cal O}_{3} \; d\sigma_{3}^{(2)}.
\end{eqnarray}
The computation of the NNLO correction for the process $e^+ e^- \rightarrow \mbox{3 jets}$ 
requires the knowledge of the amplitudes
for the three-parton final state 
$e^+ e^- \rightarrow \bar{q} q g$ up to two-loops \cite{Garland:2001tf,Moch:2002hm},
the amplitudes of the four-parton final states
$e^+ e^- \rightarrow \bar{q} q g g$
and
$e^+ e^- \rightarrow \bar{q} q \bar{q}' q'$
up to one-loop \cite{Bern:1997ka,Bern:1997sc,Campbell:1997tv,Glover:1997eh}
and the five-parton final states
$e^+ e^- \rightarrow \bar{q} q g g g$
and
$e^+ e^- \rightarrow \bar{q} q \bar{q}' q' g$
at tree level \cite{Berends:1989yn,Hagiwara:1989pp,Nagy:1998bb}.
The most complicated amplitude is of course the two-loop amplitude. For the calculation of the two-loop
amplitude special integration techniques have been 
invented \cite{Gehrmann:1999as,Moch:2001zr,Weinzierl:2002hv,Moch:2005uc}. 
The analytic result can be expressed in terms of multiple polylogarithms, which in turn
requires routines for the numerical evaluation of these 
functions \cite{Gehrmann:2001pz,Vollinga:2004sn}.

\section{Subtraction and slicing}

Is is well known that the individual pieces in the NLO and in the NNLO contribution of 
eq.~(\ref{weinzierl_nnlo_def_LO_NLO_NNLO}) are infrared divergent.
To render them finite, a mixture of subtraction and slicing is employed.
The NNLO contribution is written 
as \cite{Weinzierl:2003fx}
\begin{eqnarray}
\label{weinzierl_nnlo_nnlo_subtraction}
\langle {\cal O} \rangle^{NNLO} & = &
 \int \left( {\cal O}_{5} \; d\sigma_{5}^{(0)} 
             - {\cal O}_{4} \circ d\alpha^{single}_{4}
             - {\cal O}_{3} \circ d\alpha^{(0,2)}_{3} 
      \right) \nonumber \\
& &
 + \int \left( {\cal O}_{4} \; d\sigma_{4}^{(1)} 
               + {\cal O}_{4} \circ d\alpha^{single}_{4}
               - {\cal O}_{3} \circ d\alpha^{(1,1)}_{3}
        \right) \nonumber \\
& & 
 + \int \left( {\cal O}_{3} \; d\sigma_3^{(2)} 
               + {\cal O}_{3} \circ d\alpha^{(0,2)}_{3}
               + {\cal O}_{3} \circ d\alpha^{(1,1)}_{3}
        \right).
\end{eqnarray}
$d\alpha^{single}_{4}$ is the NLO subtraction term for $4$-parton configurations,
$d\alpha^{(0,2)}_{3}$ and $d\alpha^{(1,1)}_{3}$ are generic NNLO subtraction terms,
which can be further decomposed into
\begin{eqnarray}
 d\alpha^{(0,2)}_{3} & = & d\alpha^{double}_{3} + d\alpha^{almost}_{3} + d\alpha^{soft}_{3} 
                           - d\alpha^{iterated}_{3},
 \nonumber \\
 d\alpha^{(1,1)}_{3} & = & d\alpha^{loop}_{3} + d\alpha^{product}_{3}
                         - d\alpha^{almost}_{3} - d\alpha^{soft}_{3} + d\alpha^{iterated}_{3}.
\end{eqnarray}
In a hybrid scheme of subtraction and slicing the subtraction terms have to satisfy weaker
conditions as compared to a strict subtraction scheme.
It is just required that 
\begin{description}
\item{(a)} the explicit poles in the dimensional regularisation parameter $\varepsilon$
in the second line of eq.~(\ref{weinzierl_nnlo_nnlo_subtraction})
cancel after integration over unresolved phase spaces
for each point of the resolved phase space.
\item{(b)} the phase space singularities 
in the first and in the second line of eq.~(\ref{weinzierl_nnlo_nnlo_subtraction}) 
cancel after azimuthal averaging has been performed.
\end{description}
Point (b) allows the determination of the subtraction terms from spin-averaged matrix elements.
The subtraction terms can be found 
in \cite{Gehrmann-DeRidder:2004tv,GehrmannDeRidder:2007jk,Weinzierl:2006ij}.
The subtraction term $d\alpha^{(0,2)}_{3}$ without $d\alpha^{soft}_{3}$ would approximate 
all singularities except a soft single unresolved singularity. The subtraction term
$d\alpha^{soft}_{3}$ takes care of this last piece \cite{Weinzierl:2008iv,Weinzierl:2009nz}.
The azimuthal average is not performed in the Monte Carlo integration.
Instead a slicing parameter $\eta$ is introduced to regulate the phase space singularities
related to spin-dependent terms.
It is important to note that there are no numerically large contributions proportional
to a power of $\ln \eta$ which cancel between the 5-, 4- or 3-parton contributions.
Each contribution itself is independent of $\eta$ in the limit $\eta\rightarrow 0$.

\section{Monte Carlo integration}

The integration over the phase space is performed numerically with Monte Carlo techniques.
Efficiency of the Monte Carlo integration is an important issue, especially for the first moments
of the event shape observables. Some of these moments receive sizable contributions from the close-to-two-jet region.
In the 5-parton configuration this corresponds to (almost) three unresolved partons.
The generation of the phase space is done sequentially, starting from a 2-parton configuration.
In each step an additional particle is inserted \cite{Weinzierl:2006ij,Weinzierl:2006yt}.
In going from $n$ partons to $n+1$ partons, the $n+1$-parton phase space is partitioned into different channels.
Within one channel, the phase space is generated iteratively according to
\begin{eqnarray}
 d\phi_{n+1} & = & d\phi_n d\phi_{unresolved\;i,j,k}
\end{eqnarray}
The indices $i$, $j$ and $k$ indicate that the new particle $j$ is inserted between the hard radiators $i$ and $k$. 
For each channel we require that the product of invariants $s_{ij} s_{jk}$ is the smallest among all considered channels.
For the unresolved phase space measure we have
\begin{eqnarray}
 d\phi_{unresolved\;i,j,k}
 & = & \frac{s_{ijk}}{32 \pi^3} 
       \int\limits_0^1 dx_1 
       \int\limits_0^1 dx_2  
       \int\limits_0^{2\pi} d\varphi \; \Theta(1-x_1-x_2)
\end{eqnarray}
We are not interested in generating invariants smaller than $(\eta s)$, these configurations will be rejected by the slicing procedure.
Instead we are interested in generating invariants with values larger than $(\eta s)$ with a distribution which 
mimics the one of a typical matrix element.
We therefore generate the $(n+1)$-parton configuration from the $n$-parton configuration by using three random numbers
$u_1$, $u_2$, $u_3$ uniformly distributed in $[0,1]$ and by setting
\begin{eqnarray}
 x_1 = \eta_{PS}^{u_1}, \;\;\; x_2 = \eta_{PS}^{u_2} \;\;\; \varphi = 2 \pi u_3.
\end{eqnarray}
The phase space parameter $\eta_{PS}$ is an adjustable parameter of the order of the slicing parameter $\eta$.
The invariants are defined as
\begin{eqnarray}
 s_{ij} = x_1 s_{ijk},
 \;\;\;
 s_{jk} = x_2 s_{ijk},
 \;\;\;
 s_{ik} = (1-x_1-x_2) s_{ijk}. 
\end{eqnarray}
From these invariants and the value of $\varphi$ we can reconstruct the four-momenta of the $(n+1)$-parton configuration
\cite{Weinzierl:1999yf}.
The additional phase space weight due to the insertion of the $(n+1)$-th particle is
\begin{eqnarray}
 w & = & \frac{1}{16 \pi^2} \frac{s_{ij} s_{jk}}{s_{ijk}} \ln^2\eta_{PS}.
\end{eqnarray}
Note that the phase space weight compensates the typical eikonal factor $s_{ijk}/(s_{ij}s_{jk})$ of a single emission.
As mentioned above, the full phase space is constructed iteratively from these single emissions.

\section{Numerical results}

\begin{figure}
\begin{center}
\includegraphics[bb= 125 460 490 710,width=0.48\textwidth]{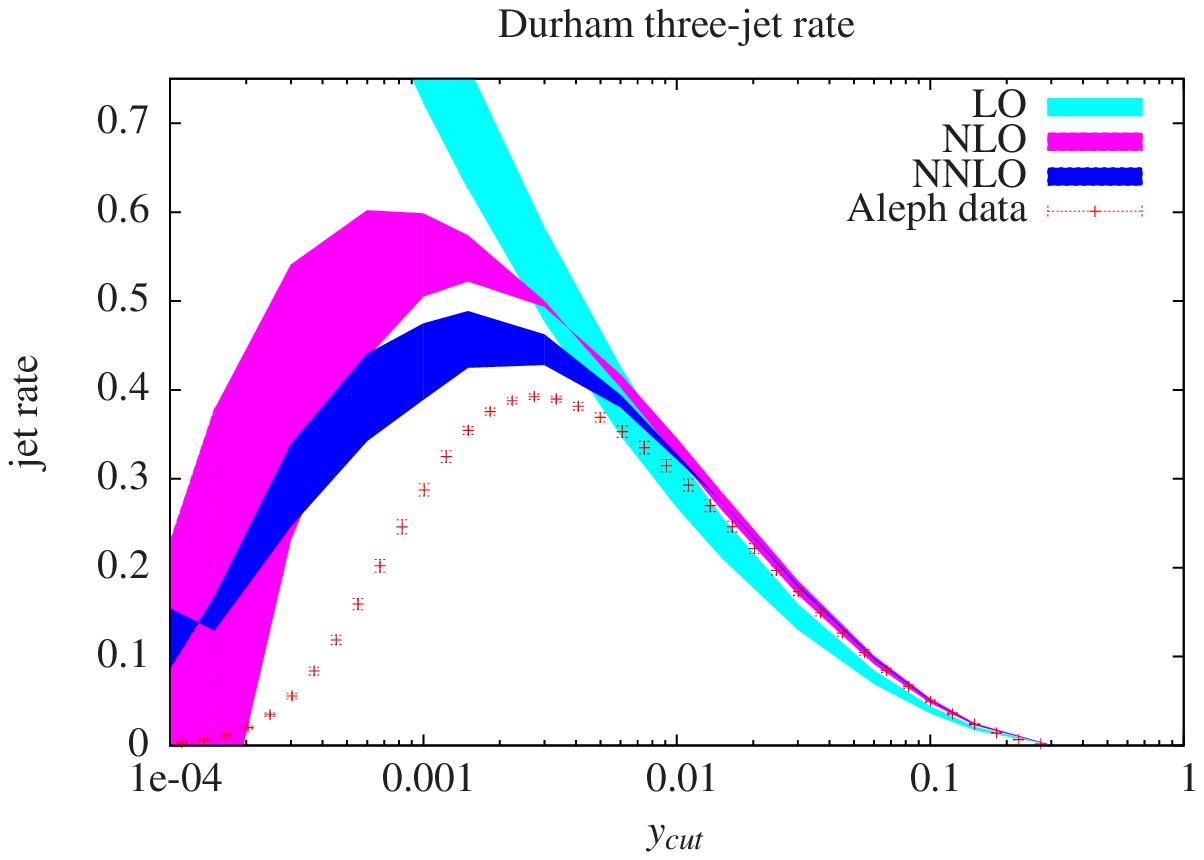}
\includegraphics[bb= 125 460 490 710,width=0.48\textwidth]{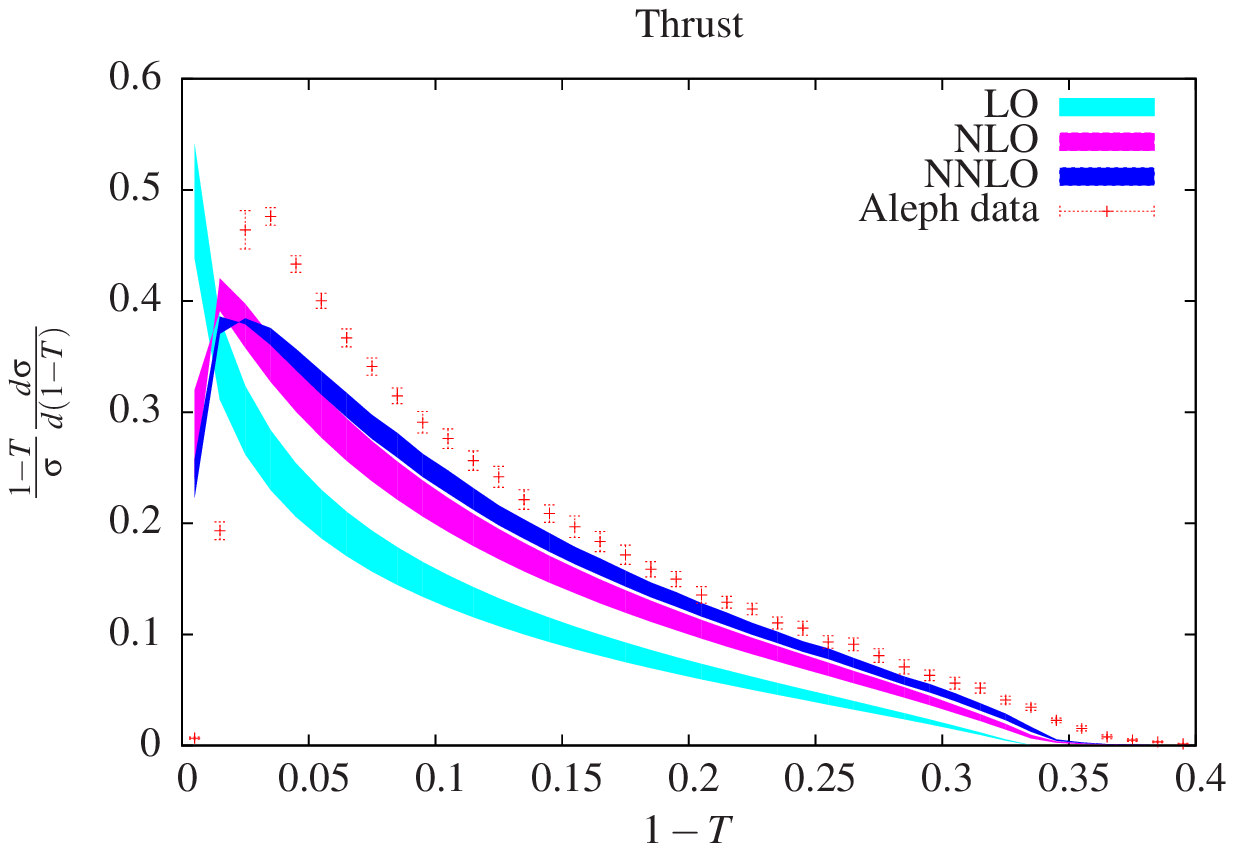}
\end{center}
\caption{The scale variation of the Durham three jet rate and the thrust distribution at $\sqrt{Q^2}=m_Z$ with $\alpha_s(m_Z)=0.118$.
The bands give the range for the theoretical prediction obtained from varying the renormalisation scale
from $\mu=m_Z/2$ to $\mu=2 m_Z$.
}
\label{fig_band}
\end{figure}

Fig.~\ref{fig_band} shows the results for the Durham three jet rate and the thrust distribution 
at the LEP I centre-of-mass energy $\sqrt{Q^2}=m_Z$ with $\alpha_s(m_Z)=0.118$.
The LO, NLO and NNLO predictions are shown together with the experimental measured values from the Aleph experiment \cite{Heister:2003aj}.
The bands give the range for the theoretical prediction obtained from varying the renormalisation scale
from $\mu=Q/2$ to $\mu=2 Q$.
Note that the theory predictions in these plots are the pure perturbative predictions. Power corrections or soft gluon resummation
effects are not included in these results.

\begin{figure}
\begin{center}
\includegraphics[bb= 125 460 490 710,width=0.32\textwidth]{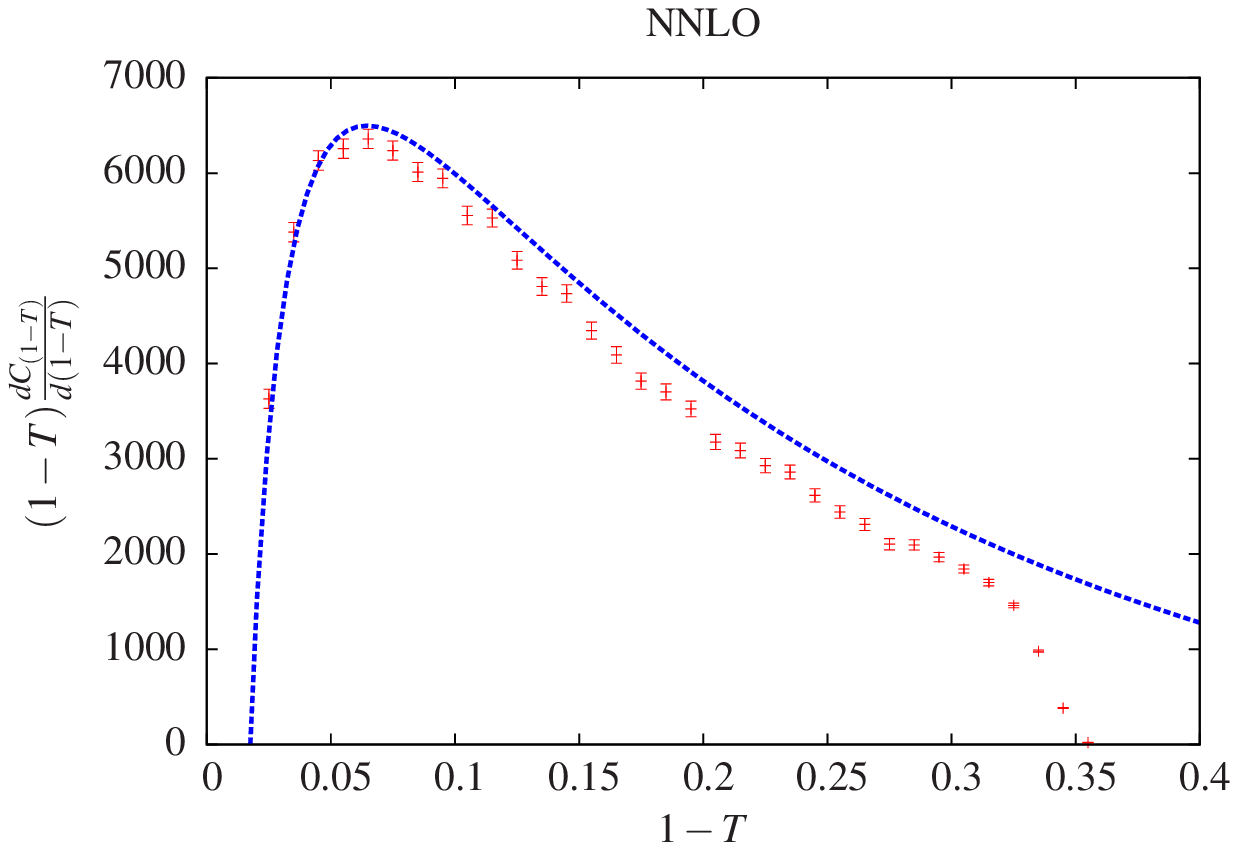}
\includegraphics[bb= 125 460 490 710,width=0.32\textwidth]{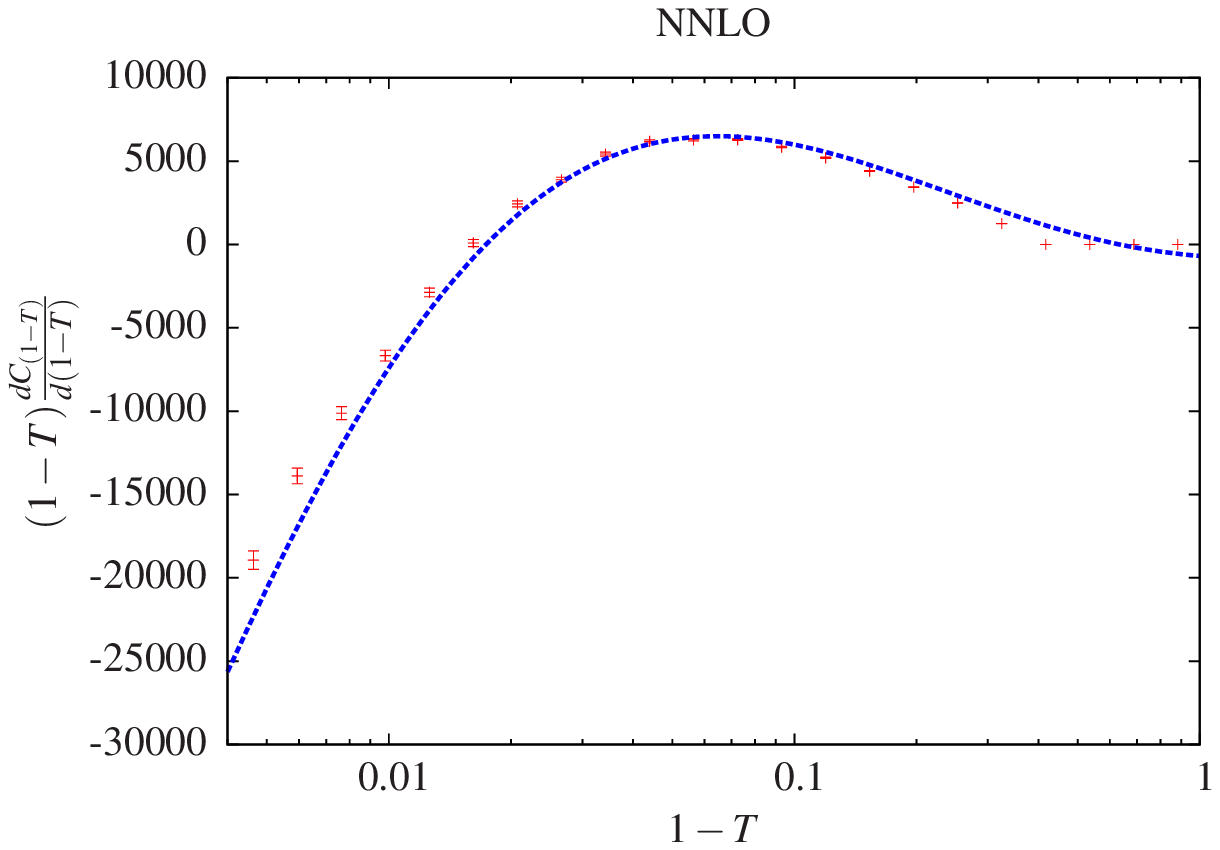}
\includegraphics[bb= 125 460 490 710,width=0.32\textwidth]{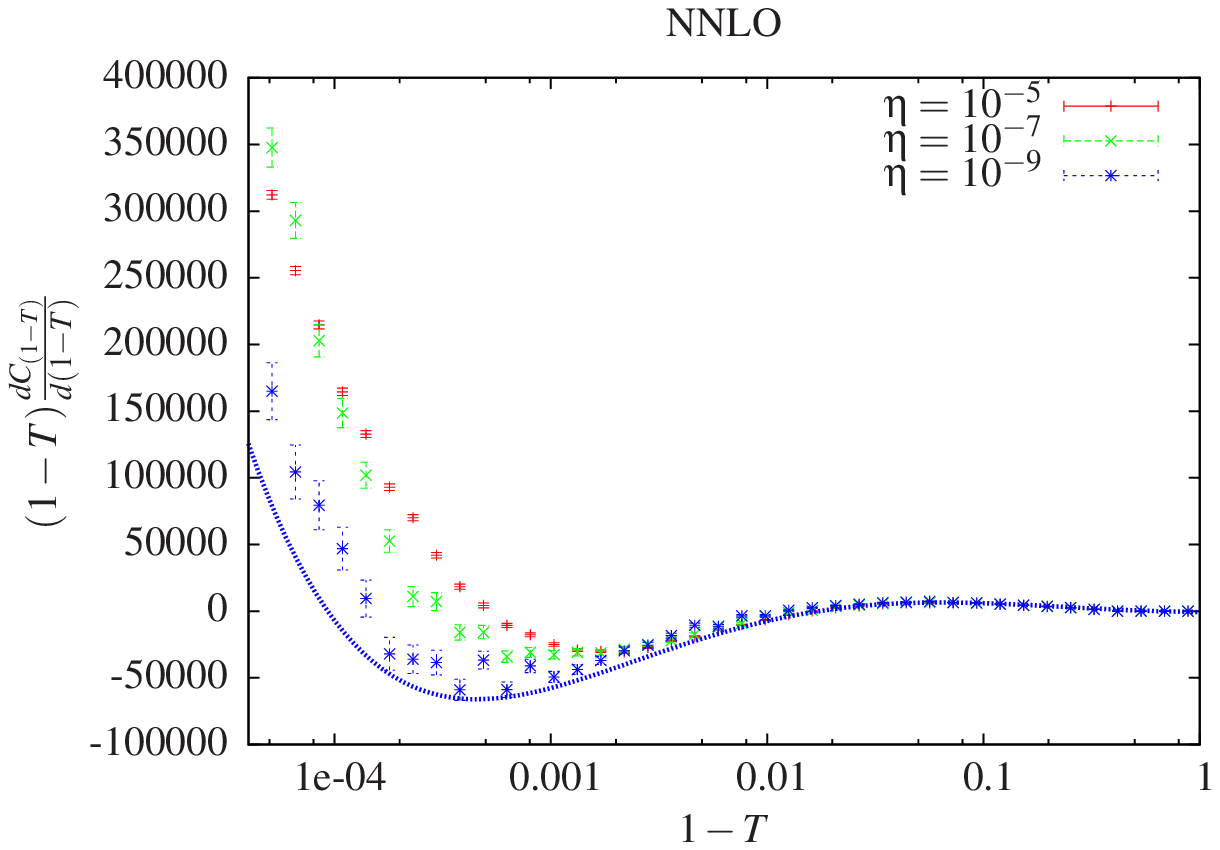}
\end{center}
\caption{A comparison of the NNLO coefficient of the thrust distribution as obtained from the numerical program
with the logarithmic terms obtained from SCET.
}
\label{fig_thrust}
\end{figure}

In a recent calculation the logarithmic terms of the NNLO coefficient of the thrust distribution have been 
calculated based on soft-collinear effective theory \cite{Becher:2008cf}:
\begin{eqnarray}
\label{log_terms}
 \frac{dC_{\tau}}{d\tau}
 & = &
 \frac{1}{\tau} \left[ a_5 \ln^5\tau + a_4 \ln^4\tau + a_3 \ln^3\tau + a_2 \ln^2\tau + a_1 \ln\tau + a_0 + {\cal O}(\tau) \right],
 \;\;\;
 \tau = 1-T.
\end{eqnarray}
The values of the $a_j$'s are for $N_f=5$
\begin{eqnarray*}
& &
a_5 = -18.96,
 \;\;\;
a_4 = -207.4,
 \;\;\;
a_3 = -122.3,
 \;\;\;
a_2 = 1488.3,
 \;\;\;
a_1 = -822.3,
 \;\;\;
a_0 = -683.4.
\end{eqnarray*}
The logarithmic terms give a good description of the thrust distribution in the close-to-two jet region.
They are not expected to give an accurate result in the hard region.
Fig.~\ref{fig_thrust} shows the comparison of the NNLO coefficient of the thrust distribution as obtained from the
numerical program with eq.~(\ref{log_terms}).
In the left plot of fig.~\ref{fig_thrust} the x-axis shows $(1-T)$ on a linear scale. This corresponds to the hard
region, where the NNLO result from the numerical program is expected to give the correct answer.
The middle plot of fig.~\ref{fig_thrust} shows $(1-T)$ on a logarithmic scale around $(1-T) \approx 0.1$. This 
corresponds to the peak region or the overlap region, where the perturbative NNLO result and the one obtained from SCET 
agree.
The right plot of fig.~\ref{fig_thrust} shows $(1-T)$ on a logarithmic scale around $(1-T) \approx 0.001$. This 
corresponds to the extreme two-jet region, in which the logarithmic terms are dominant.
In this region the results from the numerical program show a dependence on the slicing parameter.
The numerical results for $\eta=10^{-5}$, $\eta=10^{-7}$ and $\eta=10^{-9}$ are plotted.
For smaller values of $\eta$ the SCET result is approached.



\end{document}